\documentstyle[pre,aps,psfig,twocolumn]{revtex}
\begin{document}
\draft
\flushbottom
\twocolumn[
\hsize\textwidth\columnwidth\hsize\csname @twocolumnfalse\endcsname

\title{Closed almost-periodic orbits in semiclassical quantization
of generic polygons}
\author{Debabrata Biswas}
\address{Theoretical Physics Division, Bhabha Atomic Research Centre,
Trombay, Mumbai 400 085, India }

\maketitle

\tightenlines
\widetext
\advance\leftskip by 57pt
\advance\rightskip by 57pt

\begin{abstract}

Periodic orbits are the central ingredients of modern semiclassical
theories and corrections to these are generally non-classical in
origin. We show here that for the class of generic polygonal
billiards, the corrections are predominantly classical in origin
owing to the contributions from closed almost-periodic (CAP) orbit 
families. Furthermore, CAP orbit families outnumber periodic families
but have comparable weights. They are hence indispensable for 
semiclassical quantization.

\end{abstract}

\date{today}
]
\narrowtext
\tightenlines

\newcommand{\be}{\begin{equation}}
\newcommand{\ee}{\end{equation}}
\newcommand{\bea}{\begin{eqnarray}}
\newcommand{\eea}{\end{eqnarray}}
\newcommand{\Lop}{{\cal L}}
\newcommand{\DB}[1]{\marginpar{\footnotesize DB: #1}}

There exists an approximate dual relationship between the spectrum of
quantum energy eigenvalues and the classical length spectrum
of periodic orbits and this forms the central theme of modern
semiclassical theories. This duality was first discovered for the case of
hyperbolic dynamics where all periodic orbits are isolated and
unstable \cite{gutzwiller} and it was subsequently extended to the
case of marginally stable systems where
periodic orbits occur in families \cite{berry76}.
In particular, within the class of
billiard systems (particle moving freely inside an enclosure and
reflecting specularly from the walls),
such a duality exists for polygons which are marginally stable
and where periodic orbits with even bounces occur in bands \cite{pjr}.

In general, there are other (weaker) {\em non-classical} contributions
that make the relationship only approximate
\cite{diffraction} and must be included at finite energy.
For special cases however (the tilted stadium billiard \cite{harel_uzy}
and the truncated hyperbola billiard \cite{aurich_etal})
there is a source of classical correction as well.
The aim of this paper is to show that for an entire class of systems,
corrections to the periodic orbit sum are predominantly
{\em classical} in origin and are due to closed almost-periodic
orbits. Also, because they are more numerous and have weights
comparable to those of periodic orbit families, such orbits
are indispensable at finite energies.
First, however, we shall outline the key steps leading to the
{\em semiclassical trace formula} where periodic orbits are the
sole classical ingredients.

A convenient starting point is the relation \cite{gutzwiller}

\bea
\sum_n {1\over E - E_n} & = & \int~dq~G(q, q; E) \\
                        & \simeq & \int~dq~G_{s.c.}(q, q; E) \label{eq:basic}
\eea

\noindent
where $G$ and $G_{s.c.}$ refer respectively to the exact and
semiclassical
energy dependent propagator (Green's function) and $\{E_n\}$ are the
energy eigenvalues. The approximate propagator,
$G_{s.c.}$ is obtained from a fourier
transform of the semiclassical time dependent propagator \cite{gutzwiller}
and for a billiard,

\be
G_{s.c}(q, q'; E) = - \imath \sum {1\over \sqrt{8\pi \imath k l(q,q')}}
e^{\imath k l(q,q')  - \imath \mu \pi/2} \label{eq:G_sem}
\ee

\noindent
where the sum runs over all orbits at energy $E = k^2$ between $q$ and $q'$
having length $l(q,q')$ and $\mu$ is the associated Maslov
index. For convenience,
we have chosen the mass $m = 1/2$ and $\hbar = 1$.

In the limit $k \rightarrow \infty$, the amplitude
term in eq.~(\ref{eq:G_sem}) varies slowly and can be regarded
as a constant. The contribution
of a particular orbit thus depends solely on the rapidity with which
its action changes as $q$ is varied.
For periodic orbits, the action $S(q,q)$ does not vary along the
orbit. Further, if it occurs in a band, the action does not
vary in the transverse direction either and the $q$-integration
merely picks up the area, $a_p$, of the primitive band. Thus

\bea
\rho(E) & = & \sum_n \delta (E - E_n) = - {1\over \pi} \lim_{\epsilon
\rightarrow 0} \Im {1\over E + i \epsilon - E_n} \nonumber \\
        & \simeq & \rho_{av}(E) + \sum_p \sum_{r=1}^{\infty} {a_p \over
\sqrt{8\pi^3 k rl_p}} \nonumber \\
& \times & \cos(k rl_p - \pi/4) -
\sum_{p'} \sum_{r'=1}^{\infty} {l_{p'} \over 4\pi k} \cos(k r'l_{p'})
 \label{eq:semi_poly}.
\eea

\noindent
where $\rho_{av}$ is the average density of states and
the sums over $p$ and $p'$ run over primitive {\em families}
and (marginally stable) {\em isolated orbits} respectively
having length $l_p$. 

For an isolated {\em unstable} periodic orbit on the other hand,
the transverse direction leads to closed orbits with actions
that vary depending on the stability of the periodic orbit and its
contribution to the trace depends on the eigenvalues of the
Jacobian matrix arising from a linearization of the transverse
flow. In contrast, closed non-periodic orbits generally have negligible
weight since their action varies rapidly with $q$. In
case of the tilted stadium \cite{harel_uzy} however, there
exists a {\em family} of closed non-periodic orbits
for which the variation of action across the family (bouncing between the
straight edges) is small and its contribution can be of the same
order as the bouncing-ball periodic orbit family in the
zero-tilt stadium. Due to its close association with orbit
families in straight-edged billiards,
it is surprising to note that diffraction
\cite{shudo_shimizu,pavloff,whelan,bruss_whelan,uzy_etal,sieber_etal}
is still considered the most significant source of correction in
generic polygonal enclosures. While this is certainly true
when the set of allowed momenta is small, generic polygons
have additional classical contributions that are by far more
important.

\begin{figure}[tbp]
{\hspace*{0.05cm}\psfig{figure=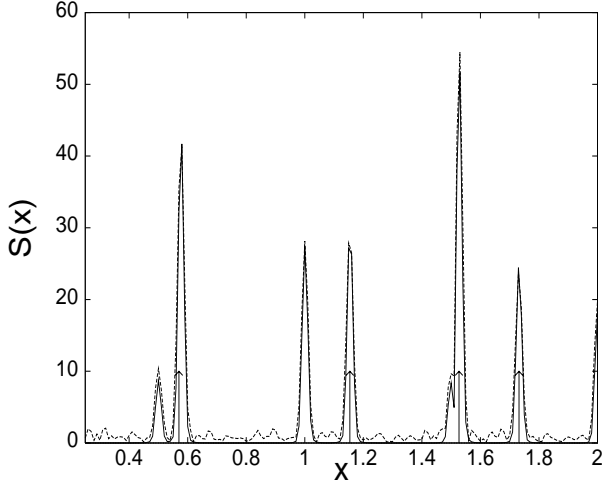,height=6.25cm,width=8cm,angle=270}}
{\vspace*{.13in}}
\caption[ty]{Length spectrum $S(x)$ of the equilateral 
and ($1001\pi/3000, 999\pi/3000$) 
triangle (referred to as $T1$)
The perimeter in both cases is 1. The arrows mark
the positions of orbits that are periodic in the equilateral triangle
but are almost-periodic in $T1$. The full and dashed lines 
correspond to the equilateral and $T1$ triangles respectively.
In both cases, the first 1100 levels have been used to obtain $S(x)$.
}
\label{fig:1}
\end{figure}

To underscore this point, consider an arbitrary polygon $T_i$
obtained by perturbing another arbitrary polygon $T$.
The slight change in the shape of the enclosure
results in a slight change in the quantal eigenenergies so
that the structure of the length spectrum, $S(x)$ (the power 
spectrum of $\rho(k) = 2k \rho(E) $), is largely preserved
and there are only minor variations in peak heights (see fig.~\ref{fig:1}).
However, the spectrum of periodic orbit lengths in $T$ and $T_i$ 
are radically different as we shall shortly demonstrate.
There is thus an apparent paradox which cannot be resolved
by invoking diffraction since their contributions are
${\cal O}(k^{-1})$ at best \cite{EB99} compared to 
the ${\cal O}(k^{-1/2})$ contributions of geometric periodic families.

The change in length spectrum of periodic orbits upon perturbation
is best illustrated by comparing the equilateral and $T1$ triangles.
As in case of all rational polygons, the invariant surface of $T1$
is two dimensional and topologically equivalent to a sphere with
$g$ holes where $g = 1 + ({\cal N}/2) \sum_i (m_i - 1)/n_i$ where
$\{m_i\pi/n_i$\} are the internal angles of the triangle
and ${\cal N}$ is the least common multiple of $\{n_i\}$. Thus
for the $T1$ triangle, $g = 1000$
while for the equilateral triangle, $g = 1$. Note that the
number of allowed momenta values is $2{\cal N}$ so that if
${\cal N}$ is large, the probability that two segments of
a trajectory have a small angle intersection is large.
Thus, even though the boundary is only slightly perturbed,
the structure of the invariant surface changes radically.
It may thus be expected that the spectrum of periodic orbit lengths
in the two systems is very different as well.
In the integrable case, these
invariant trajectories live on the torus and are labelled by the
winding numbers $(M_1,M_2)$ which count the number of times the
orbit goes around the two irreducible circuits. In the non-integrable
case, very little prior information is available \cite{gutkin}
and we shall analyze the situation to
demonstrate that the symbol
sequences of periodic orbits in the equilateral triangle do not
necessarily lead to periodic orbits in $T1$.

For the triangle enclosures, we shall use the symbols
$\{1,2,3\}$ for the three sides \cite{binary}. A trajectory
can then be labelled by a string of symbols $s_1s_2\ldots s_n$
where $s_i \in \{1,2,3\}$. Thus a sequence $1323$
denotes a trajectory that reflects off sides 1, 3, 2 and 3
respectively. Let us denote by $R_i, i=1,3$ the $2 \times 2$
reflection matrices of the three sides. These can be
expressed in terms of the angle $\theta_i$ between the outward normal
(  ${\hat n}_i$ ) to a side and the positive $X$-axis :

\be
R_i = \left ( \matrix{ -\cos(2\theta_i) & -\sin(2\theta_i) \cr
               -\sin(2\theta_i) &\;\;\cos(2\theta_i) } \right ) .
\label{eq:ref_matrix}
\ee

\noindent
Thus, for the sequence $1323$, the initial and final
velocities are related by

\be
\left ( \begin{array}{c} v_x^f \\ v_y^f \end{array} \right ) =
R_3 \circ R_2 \circ R_3 \circ R_1
\left ( \begin{array}{c} v_x^i \\ v_y^i \end{array} \right )
= R_{1323} \left ( \begin{array}{c} v_x^i \\ v_y^i \end{array} \right )
\ee

\noindent
where the superscripts $f(i)$ refer respectively to final (initial)
velocities $\vec{v}$ whose components are $v_x$ and $v_y$.
It is easy to verify that when the number of reflections
is odd

\be
R^{(odd)}_{s_1s_2 \ldots s_n} =
\left ( \matrix{ -\cos(\varphi_o) & -\sin(\varphi_o) \cr
                 -\sin(\varphi_o) &\;\; \cos(\varphi_o) } \right )
\ee

\noindent
where $\varphi_o = 2 (\theta_1 + \theta_3 + \ldots + \theta_n) -
2(\theta_2 + \theta_4 + \ldots + \theta_{n-1})$ while for
even number of reflections ($n$ even)

\be
R^{(even)}_{s_1s_2 \ldots s_n} =
\left ( \matrix{ \;\;\cos(\varphi_e) & \sin(\varphi_e) \cr
                 -\sin(\varphi_e) & \cos(\varphi_e) } \right )
\ee

\noindent
where $\varphi_e = 2 (\theta_1 + \theta_3 + \ldots + \theta_{n-1}) -
2(\theta_2 + \theta_4 + \ldots + \theta_{n})$.

Obviously, the initial
and final velocities can be equal if the resultant reflection matrix
$R_{s_1s_2\ldots s_n}$ has a unit eigenvalue.
For even $n$ (the case of bands or families), the eigenvalues
are $e^{\pm \imath \varphi_e}$ so that the condition for the
existence of a unit eigenvalue is

\be
\varphi_e = 0 \hspace{0.1in} {\rm mod}(2\pi).
\label{eq:even_condition}
\ee

\noindent
For odd $n$ on the other hand, the product of the eigenvalues
$\lambda_1 \lambda_2 = 1$.
The eigenvector corresponding to a unit
eigenvalue is ($\sin(\varphi_o/2) , -\cos(\varphi_o/2)$)
so that if a real
orbit exists with the sequence $s_1s_2\ldots s_n$, its initial
and final velocities are equal.

In the event that a sequence repeats itself (denoted by
$\overline{s_1s_2\ldots s_n}$)
and there exists a unit eigenvalue of the resultant matrix
$R_{s_1s_2\ldots s_n}$, stability
considerations guarantee that a periodic orbit exists \cite{DB99_3}.
When $n$ is odd, the orbit is isolated where as when $n$ is
even the orbit exists in an equi-action family.

Not all sequences are however allowed.
Further, not all repeating sequences guarantee the existence of
periodic orbits due to eq.~(\ref{eq:even_condition}). For the
$T1$ triangle, it is clear that
the set of repeating sequence are the same as in the
equilateral triangle for short orbits. Eq.~(\ref{eq:even_condition}) however
does not allow all of them to be periodic. For instance,
the sequence $1323$ results in a bouncing ball family of
periodic orbits in the equilateral triangle. In the
$T1$ triangle however, the eigenvalues for this
sequence are $\exp(\pm \imath\pi/1500)$
so that there can be no periodic orbit with  reflections
from these sides. A sequence that is however allowed
and leads to periodic orbit families in both triangles is
$123123$ ( this is distinct from $\overline{123}$)
since the periodicity condition (eq.~\ref{eq:even_condition})
is automatically satisfied. In general then, {\em for an arbitrary
enclosure close to the equilateral triangle, an allowed
sequence that repeats itself in the equilateral case can
be a periodic family only when each symbol occurs as many times in
even places as in the odd places}. Thus, corresponding to the sequence
$3231231231$, there does not exist any periodic orbit in the $T1$ triangle
while a periodic family exists in the equilateral case.

We have thus verified that the periodic orbits
in the $T1$ and  equilateral
triangles are indeed different even though short orbits
follow the same sequence due to the proximity of the two
triangles. Note that this observation holds in general
for any arbitrary enclosure $T$. Upon perturbation, orbits
follow the same sequence but the periodicity condition
will not be satisfied for sequences that are periodic in $T$.
According to eq.~(\ref{eq:semi_poly}) therefore,
the peak positions and heights in the length spectrum
should differ and we shall now show 
that the similarity in length spectrum observed in fig. \ref{fig:1} 
is due to contributions from closed almost-periodic orbit
families in $T1$.

\begin{figure}[tbp]
{\hspace*{0.5cm}\psfig{figure=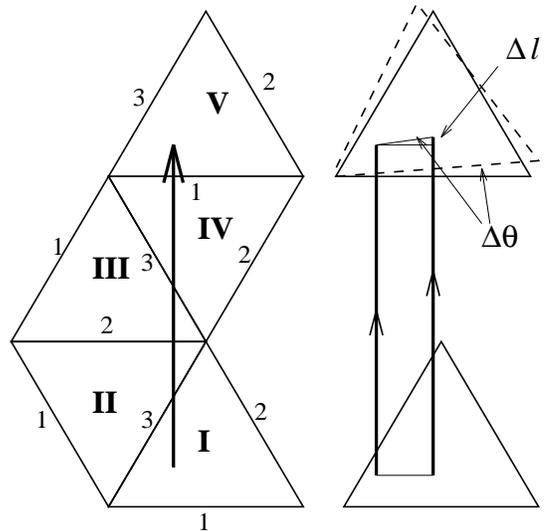,height=7cm,width=7cm,angle=270}}
{\vspace*{.13in}}
\caption[ty]{The unfolded trajectory $\overline{3231}$ (marked by
an arrow) is produced by successive reflections of triangle I to
produce copies II, III, IV and V. For the equilateral case, copy
I and V have the same orientation and the trajectory is periodic.
For $T1$, the orientations
differ slightly as shown schematically in the right. As a result
the orbits are closed but non-periodic.}
\label{fig:2}
\end{figure}

Consider a symbol sequence that repeats itself and exists in
both the equilateral and the $T1$ triangles. Further,
assume that corresponding to this sequence, there does not
exist any periodic
orbit in the $T1$ triangle while a periodic orbit family does exist
in the equilateral case. Examples of these are the sequences
$\overline{3231}$,  $\overline{3231231231}$ ($l_p = 1.5275$) and
$\overline{2312312312312131}$ $ (l_p = 2.5166) $.
In every such case, one can construct  ``unfolded'' trajectories
(which are straight lines) by successive reflections of the
triangle about the sides where the collision occurs.
For instance (see fig. \ref{fig:2} ),
unfolded trajectories for the sequence $3231$ can be created by
first reflecting the triangle about side 3. The copy (II) so obtained is
then reflected about side 2, the resultant copy (III) reflected about
side 3 and finally (copy IV) about side 1. For the equilateral triangle,
the final copy (V) has the same orientation as the initial copy (I)
so that any line joining corresponding points in the initial
and final copies is an ``unfolded  periodic orbit''. In the
$T1$ triangle however, the final copy differs marginally
in orientation from the initial copy so that any line joining
corresponding points in the two can only be a closed almost-periodic
orbit. Obviously, at every point ${\bf q}$ there exists such a closed orbit
with this sequence so long as the line joining the corresponding
points (in I and V) lies entirely within the copies
generated through reflections.
Two such orbits separated by $q_\bot$ are shown in fig. \ref{fig:2} (right).
It is easy to see that the orbits differ in length by an amount
$\Delta l =  q_\bot \tan(\Delta \theta) \simeq q_\bot~
\Delta \theta$ if $\Delta \theta$ is small.
Note that the above analysis holds for other almost-periodic closed
orbits as well (such as the sequence $3231231231$)
and any arbitrary polygon, $T$. In each of these
cases $\Delta l \simeq
  q_\bot~\Delta \theta = q_\bot~ \varphi_e$
so that the length varies slowly if the orbit nearly closes in momentum. 

The correct trace formula for an arbitrary polygon $T$
can be derived by noting that for a closed almost-periodic
family, $l(q_\bot) = l(0) + q_\bot \varphi_e$
where $l(0) = l_i$ is the length of the orbit in
the centre of the band and $q_\bot$ varies from $-w_i/2$ to
$w_i/2$ where $w_i$ is the transverse extent of the band.
Assuming that $k$ is sufficiently large, the amplitude
($1/l(q_\bot)$) can be treated as a constant ($1/l_i$)
and the trace formula for finite $k$ is then

\bea
\rho(E) & \simeq & \rho_{av}(E) + \sum_i  {a_i \over
\sqrt{8\pi^3 k l_i}} \nonumber \\
& \times & \cos(kl_i  -  \pi/4)
{\sin(k\varphi_e^{(i)} w_i/2)\over k\varphi_e^{(i)} w_i/2}
 \nonumber \\
& - & \sum_{p'} \sum_{r'=1}^{\infty} {l_{p'} \over 4\pi k} \cos(k r'l_{p'}).
\label{eq:semi_poly_modif}
\eea

\noindent
In eq.~(\ref{eq:semi_poly_modif}),
the sum over $i$ runs over closed almost-periodic and periodic
orbit families and $l_i$ is the (average) length
of such a family. Note that as $k \rightarrow \infty$, the
contribution of almost-periodic orbits ($\varphi_e^{(i)} \ne 0$)
vanishes as $k^{-3/2}$ so that eq.~(\ref{eq:semi_poly_modif}) reduces to
eq.~(\ref{eq:semi_poly}). For de Broglie wavelength, $\lambda
 >> \pi w_i \varphi_e^{(i)}$, however,
the ($i$th) closed almost-periodic orbit family contributes
with a weight comparable to that of periodic families 
(${\cal O}(1/k^{1/2})$)
and hence assumes greater significance than diffraction
\cite{low_genus}.
Interestingly, such orbits clearly show up in eigenfunctions
\cite{bellomo} and this has been referred to as
``scarring by ghosts of periodic orbits'' since such a
periodic orbit exists only in a neighbouring polygon.
Thus a direct resolution of the paradox lies in closed almost-periodic
orbits.

\begin{figure}[tbp]
{\hspace*{0.1cm}\psfig{figure=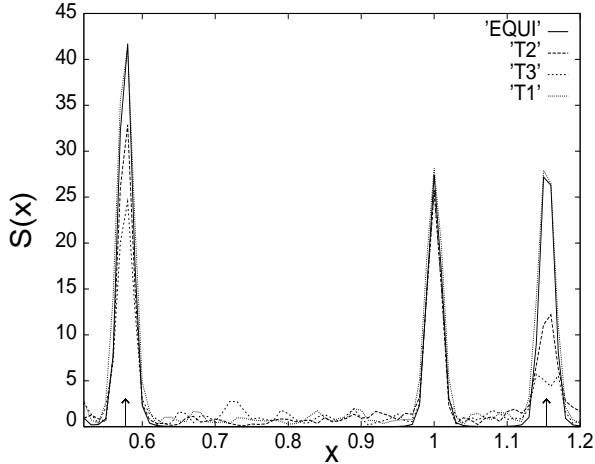,height=6cm,width=7.75cm,angle=270}}
{\vspace*{.13in}}
\caption[ty]{A comparison of the length spectrum for four
different triangles : EQUI - equilateral,
$T1$ - ($1.001\pi/3, 0.999\pi/3, \pi/3$),
$T2$ - ($1.01\pi/3, 0.99\pi/3, \pi/3$) and
$T3$ - ($1.01513\pi/3, 0.98487\pi/3, \pi/3$).
The arrows are at $0.577$ and $1.154$ corresponding to the sequence $3231$.
In all cases, the first 1100 levels have been used to obtain $S(x)$.
Note that  $T1$ is practically indistinguishable from the equilateral
curve while $T3$ is farthest from EQUI. The corresponding values
of $\varphi_e$ for the four cases are : EQUI - 0,
$T1$ - $0.000667\pi$, $T2$ - $0.006667\pi$ and
$T3$ - $0.010087\pi$. In contrast, the peak
at $x = 1$ remains
unchanged for all 4 triangles since it corresponds to a periodic
orbit ($\overline{123123}$). }
\label{fig:3}
\end{figure}

To emphasize the importance of the angle between the initial
and final momentum ($\varphi_e$), we compare the power spectrum
of three different triangles, $T1$,$T2$ and $T3$ with the
equilateral triangle in figure \ref{fig:3}. For the sequence
$3231$,  $\varphi_e$ is maximum for $T3$ and minimum for $T1$
so that peak heights at $0.57$ and its repetitions should be
closest to those of the equilateral triangle for $T1$ and
farthest for $T3$. This can indeed be verified from fig.~\ref{fig:3}.

\begin{figure}[tbp]
{\hspace*{0.1cm}\psfig{figure=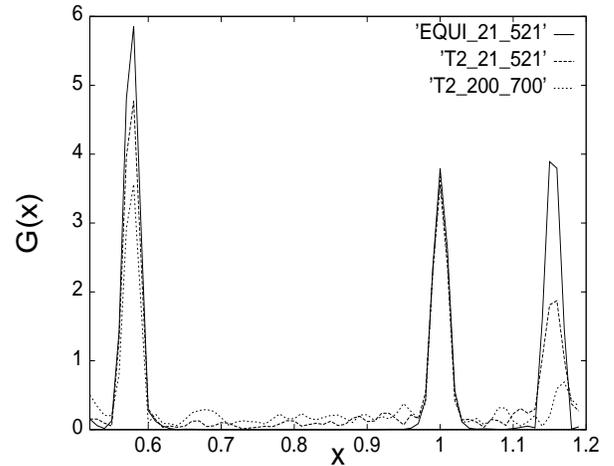,height=6cm,width=7.75cm,angle=270}}
{\vspace*{.13in}}
\caption[ty]{A comparison of $G(x)$ for 2 energy ranges for the
$T2$ triangle together with a typical plot for the equilateral
triangle (marked EQUI\_21\_521 with $k_\alpha = 21$  and
$k_\beta = 521$) when $k_{\beta} - k_{\alpha} = 500$.
Note the diminishing peak heights for the range $k \in (200,700)$.}
\label{fig:4}
\end{figure}

The contributions of CAP families diminish with energy in 
accordance with eq.~(\ref{eq:semi_poly_modif}) and can be
observed in the length spectrum. In order to distinguish this
from the contribution of periodic families, we shall consider
the power spectrum, G(x), of $\rho(k)/k^{1/2}$

\be
G(x) = \left | \sum_{k_{\alpha} \leq k_n \leq k_{\beta}} 
{\cos(k_n x) \over k_n^{1/2}} + 
\imath \sum_{k_{\alpha} \leq k_n \leq k_{\beta}}
{\sin(k_n x) \over k_n^{1/2}} \right |
\ee

\noindent
such that for a fixed $k_{\beta} - k_{\alpha}$, the peak height
of periodic families remains unaltered irrespective of $k_\beta$. 
Fig.~\ref{fig:4} show
plots of $G(x)$ for the $T2$ triangle using two different
$k$-intervals : (21,521) and (200,700). In both cases, the
peak height remains unaltered at $x = 1.0$ corresponding
to a periodic family. The peak at $x = 0.57$ however
diminishes in height as the interval shifts to a higher
energy. Also shown is a plot for the equilateral triangle
which remains unchanged so long as $k_{\beta} - k_{\alpha}$ is
fixed.

Precise checks (without using any window function) between 
the observed and expected peak height at $x = 0.57$
show that the value expected from eq.~(\ref{eq:semi_poly_modif})
is $11.3$ while the observed height is $9.6$. Undoubtedly,
there are other sources of corrections but the dominant
contribution at this value of $x$ is due to the closed
almost-periodic familiy.

Finally, though the examples chosen are close to the ($\pi/3,\pi/3$)
triangle, we wish to reiterate that closed almost-periodic families
contribute away from the neighbourhood of integrable enclosures as well.
To see this, consider an arbitrary triangle $T$. In its
immediate neighbourhood, there exists an infinity of triangles
$\{T^{(i)}\}$, {\em each with a distinct
periodic orbit spectrum} but having the same symbol sequence for
short trajectories.
Assume now that there exists a periodic orbit corresponding to the
sequence $S_k$ for the triangle $T^{(j)}$. Then, for all other
triangles in its neighbourhood, this sequence contributes
an amount (nearly) equal to the periodic orbit contribution of  $T^{(j)}$
provided $\pi w_i \varphi_e^{(i)}  << \lambda$. Thus
corresponding to {\em every}
periodic family in {\em each} of the triangles $\{T^{(i)}\}$, there
exists an almost-periodic family in the triangle $T$ whose
contribution is comparable to that of periodic orbit families in
these neighbouring triangles.

To conclude, we have demonstrated that closed almost-periodic orbit
families are more numerous and have weights comparable to that
of periodic families in polygonal billiards. They are thus indispensable
for the semiclassical quantization of generic polygons.

%\end{document}

\newcommand{\PR}[1]{{Phys.\ Rep.}\/ {\bf #1}}
\newcommand{\PRL}[1]{{Phys.\ Rev.\ Lett.}\/ {\bf #1}}
\newcommand{\PRA}[1]{{Phys.\ Rev.\ A}\/ {\bf #1}}
\newcommand{\PRB}[1]{{Phys.\ Rev.\ B}\/ {\bf #1}}
\newcommand{\PRD}[1]{{Phys.\ Rev.\ D}\/ {\bf #1}}
\newcommand{\PRE}[1]{{Phys.\ Rev.\ E}\/ {\bf #1}}
\newcommand{\JPA}[1]{{J.\ Phys.\ A}\/ {\bf #1}}
\newcommand{\JPB}[1]{{J.\ Phys.\ B}\/ {\bf #1}}
\newcommand{\JCP}[1]{{J.\ Chem.\ Phys.}\/ {\bf #1}}
\newcommand{\JPC}[1]{{J.\ Phys.\ Chem.}\/ {\bf #1}}
\newcommand{\JMP}[1]{{J.\ Math.\ Phys.}\/ {\bf #1}}
\newcommand{\JSP}[1]{{J.\ Stat.\ Phys.}\/ {\bf #1}}
\newcommand{\AP}[1]{{Ann.\ Phys.}\/ {\bf #1}}
\newcommand{\PLB}[1]{{Phys.\ Lett.\ B}\/ {\bf #1}}
\newcommand{\PLA}[1]{{Phys.\ Lett.\ A}\/ {\bf #1}}
\newcommand{\PD}[1]{{Physica D}\/ {\bf #1}}
\newcommand{\NPB}[1]{{Nucl.\ Phys.\ B}\/ {\bf #1}}
\newcommand{\INCB}[1]{{Il Nuov.\ Cim.\ B}\/ {\bf #1}}
\newcommand{\JETP}[1]{{Sov.\ Phys.\ JETP}\/ {\bf #1}}
\newcommand{\JETPL}[1]{{JETP Lett.\ }\/ {\bf #1}}
\newcommand{\RMS}[1]{{Russ.\ Math.\ Surv.}\/ {\bf #1}}
\newcommand{\USSR}[1]{{Math.\ USSR.\ Sb.}\/ {\bf #1}}
\newcommand{\PST}[1]{{Phys.\ Scripta T}\/ {\bf #1}}
\newcommand{\CM}[1]{{Cont.\ Math.}\/ {\bf #1}}
\newcommand{\JMPA}[1]{{J.\ Math.\ Pure Appl.}\/ {\bf #1}}
\newcommand{\CMP}[1]{{Comm.\ Math.\ Phys.}\/ {\bf #1}}
\newcommand{\PRS}[1]{{Proc.\ R.\ Soc. Lond.\ A}\/ {\bf #1}}


\begin{references}
\bibitem{diffraction} Contributions from diffractive orbits
are the most relevant corrections in the present context
 \cite{gabor_creep,shudo_shimizu,pavloff,whelan,bruss_whelan,uzy_etal,sieber_etal}.
\bibitem{binary} A binary alphabet is enough to describe trajectories
in triangles though even this is cumbersome in practice since periodic
orbit families proliferate only as a power law. For methods of
finding periodic orbits, see \cite{db_rapid1,pramana_97}
\bibitem{low_genus} Note that almost-periodic closed orbit
families do not generally occur
in systems where the number of directions accessible to a trajectory
is small since the (average) angle of intersection is large.
Thus diffraction is the principal source of correction in such cases.
\end{references}
\end{document}